\documentclass[aps,twocolumn,prl,amsmath,amssymb,showpacs,notitlepage,superscriptaddress]{revtex4-1}   	
\usepackage{graphicx}				% Use pdf, png, jpg, or eps§ with pdflatex; use eps in DVI mode
\usepackage{amssymb}
\usepackage{amsmath}

\usepackage{bm}
\usepackage{bbm}

\begin{document}
%\section{}
%\subsection{}

\title{Topological characterization of classical waves: the topological origin of magnetostatic surface spin waves}

%\date{\today}

\author{Kei Yamamoto}
\email{yamamoto.kei@jaea.go.jp}
\affiliation{Advanced Science Research Center, Japan Atomic Energy Agency, Tokai 319-1195, Japan}
\affiliation{Institut f\"ur Physik, Johannes Gutenberg-Universit\"at Mainz, 55128 Mainz, Germany}
\author{Guo Chuan Thiang}
\affiliation{School of Mathematical Sciences, University of Adelaide, SA 5000, Australia}
\author{Philipp Pirro}
\affiliation{Fachbereich Physik and Landesforschungszentrum OPTIMAS, Technische Universit\"at Kaiserslautern, 67663 Kaiserslautern, Germany}
\author{Kyoung-Whan Kim}
\affiliation{Institut f\"ur Physik, Johannes Gutenberg Universit\"at Mainz, 55128 Mainz, Germany}
\affiliation{Center for Spintronics, Korea Institute of Science and Technology, Seoul 02792, Korea}
\author{Karin Everschor-Sitte}
\affiliation{Institut f\"ur Physik, Johannes Gutenberg Universit\"at Mainz, 55128 Mainz, Germany}
\author{Eiji Saitoh}
\affiliation{Department of Applied Physics, University of Tokyo, Tokyo 113-8656, Japan}
\affiliation{Institute for Materials Research, Tohoku University, Sendai 980-8577, Japan}
\affiliation{Advanced Science Research Center, Japan Atomic Energy Agency, Tokai 319-1195, Japan}

\begin{abstract}
We propose a topological characterization of Hamiltonians describing classical waves. Applying it to the magnetostatic surface spin waves that are important in spintronics applications, we settle the speculation over their topological origin. For a class of classical systems that includes spin waves driven by dipole-dipole interactions, we show that the topology is characterized by vortex lines in the Brillouin zone in such a way that the symplectic structure of Hamiltonian mechanics plays an essential role. We define winding numbers around these vortex lines and identify them to be the bulk topological invariants for a class of semimetals. Exploiting the bulk-edge correspondence appropriately reformulated for these classical waves, we predict that surface modes appear but not in a gap of the bulk frequency spectrum. This feature, consistent with the magnetostatic surface spin waves, indicates a broader realm of topological phases of matter beyond spectrally gapped ones.
\end{abstract}

\maketitle

The principle of bulk-edge correspondence is a cornerstone in the field of topological phases of matter~\cite{Kitaev2009}: at the boundary of a system whose bulk frequency spectrum is topologically nontrivial, there should appear localized edge modes with eigenfrequencies in a gap of the bulk spectrum. This principle underlies the unconventional stability of chiral edge states in quantum Hall insulators~\cite{Hatsugai1993} and Dirac surface states of topological insulators~\cite{Fu2007}, and has more recently led to predictions of edge modes in various classical systems~\cite{Raghu2008,Shindou2013,Kane2014}. The bulk system topology is usually characterized by a topological invariant defined for Hamiltonians describing spatially unbounded systems with specified symmetry operations. It dictates the existence and number of topologically protected edge modes. The corresponding hallmark of these edge states is their robustness against symmetry-preserving perturbations. 

The insensitiveness of edge states to material parameters strikes a chord in the field of magnetism. Since their discovery in 1960~\cite{Eshbach1960}, ferromagnetic spin waves known as ``magnetostatic surface waves" (MSSWs) have been a subject of various experimental and theoretical studies. These edge modes owe their intrinsic chiral structure to dipole-dipole interactions. MSSWs propagate perpendicular to the ordered magnetization regardless of the sample geometry, be it a slab~\cite{Damon1961} or a sphere~\cite{Fletcher1959}. They are known to be anomalously robust against back scatterings~\cite{Chumak2009,Mohseni2018}, hinting towards a topological origin.
%which are driven by the dipole-dipole interaction, owe their intrinsic chiral structure to the coupling of the two dynamic magnetisation components which arises for a propagation perpendicular to the static magnetization. [COMMENT by KY: Sorry, I wasn't able to understand this statement.]
The chirality and robustness render them interesting for many fundamental studies, e.g.\ for non-reciprocal transport of spin~\cite{Sander2017} and heat~\cite{An2013}. 
%Three decades ago, the study of MSSWs in coupled thin films by Gr\"unberg has greatly contributed to the understanding of interlayer exchange coupling and to the discovery of the giant magnetoresistance effect~\cite{Grunberg2007}. [COMMENT by KY: Dropped by Saitoh-sensei.]
Today, in the context of magnon spintronics~\cite{Chumak2015}, MSSWs are almost exclusively used in studies of spin-wave transport in microstructures since they offer the largest decay length of all available modes and are easily excited by the commonly used inductive microwave antennas. It is therefore of fundamental interest whether MSSWs are indeed topologically protected or not.
%In the course of all these experimental studies, the existence of such a MSSW edge mode has be verified in various systems with dipole-dipole interaction.    

\begin{figure}[tbp]
\includegraphics[width=\linewidth]{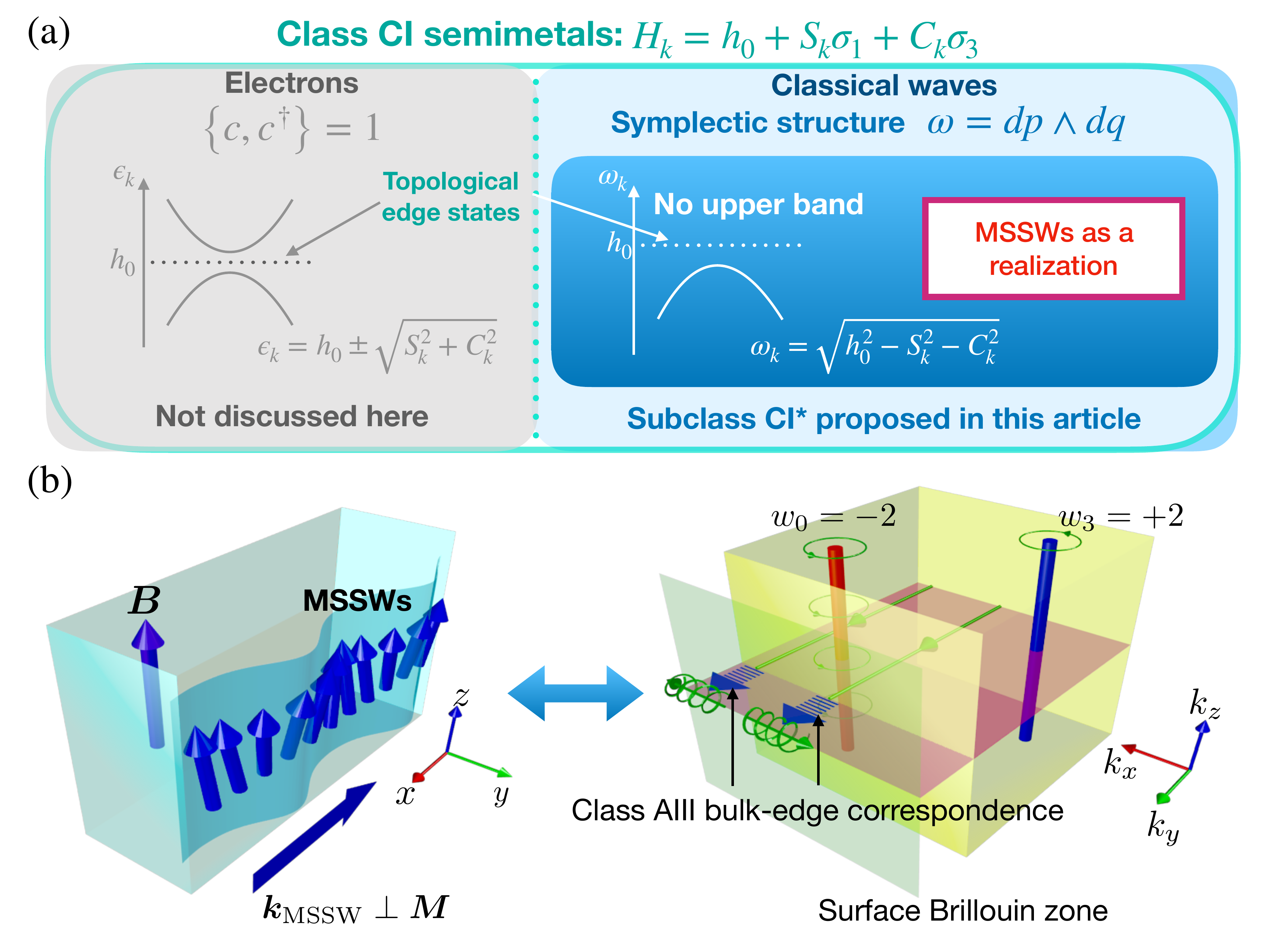}
\caption{(a) The class of systems presented in the Letter. The symmetry class is CI with $\Gamma $ and $\mathcal{T}$, but we further focus on a subclass, denoted CI*, in which $\omega $ plays the role of $\Gamma $. (b) The real space setup (Left) and the corresponding Fourier space (Right) structure for MSSWs. The green straight lines along $k_y$ axis belong to class AIII with well-defined winding numbers.}
\label{fig:geometry}
\end{figure}

In this Letter, we show that the bulk Hamiltonian of spin waves in the presence of dipole-dipole interactions is characterized by a topological invariant. A pair of \emph{vortex lines} in the Brillouin zone acts as \emph{extended Dirac monopoles}, which cannot be removed by small continuous changes in system parameters. We demonstrate that these topological vortex lines lead to MSSWs via the notion of \emph{class CI semimetals}, where CI denotes the symmetry class formally defined by the presence of two symmetry operators $\Gamma $ and $\mathcal{T} $~\cite{Ryu2010}. Even though they are conventionally called chiral and even time-reversal symmetry respectively, these mathematical operations are realized for MSSWs as the symplectic structure~\cite{Arnold1989} and the reality condition that are both inherent to classical mechanics. We first show that in a quantum mechanical context, class CI semimetals have edge states which appear in a band gap. The dipolar Hamiltonian has a topologically nontrivial class CI semimetal structure. Because it describes classical waves, however, the topological edge states have instead eigenfrequencies above the bulk spectrum, in agreement with MSSWs. Motivated by this example, we establish a new type of bulk-edge correspondence for a general class of classical mechanical systems (FIG.~\ref{fig:geometry}(a)).

%~\footnote{We note that previous studies on topological characterization of classical systems all focus on \emph{gapless} edge modes to the best of our knowledge (see \cite{Peano2018} and references therein) and are not immediately applicable to MSSWs.}

It is instructive to first visualize the setup (FIG.~\ref{fig:geometry}(b)). The 3D Brillouin zone in class CI can be sliced up into 1D subsystems (green straight lines in FIG.~\ref{fig:geometry}(b)), which generically possess only $\Gamma $-symmetry and thus belong to class AIII. As in the Su-Schrieffer-Heeger model~\cite{Su1979}, the bulk topological invariant of class AIII in 1D is the integer winding number over the 1D Brillouin zone. Its nonzero value guarantees topologically protected dangling edge modes~\cite{Ryu2002,Gomi2018} even in the presence of disorder~\cite{Prodan2016,Graf2018}. For dipolar spin waves, each subsystem gives winding number $\pm 1$ which remains constant as the slice is varied, unless a vortex line is crossed, forcing a discontinuous jump by $2$. This topological structure is analogous to Weyl semimetals, where the slice-wise 2D Chern number stays constant away from band-crossings (Weyl points) in the 3D Brillouin zone, but changes discontinuously when a Weyl point is traversed~\cite{Wan2011}. While Weyl semimetals are characterized by the Dirac monopole charges of the Weyl points (along with the Dirac strings connecting them~\cite{Mathai2017,Mathai2017a}), the dipolar spin wave Hamiltonian features vortex lines of 1D ``extended monopoles'' in 3D, i.e. topological defects of codimension two. 
%The 1D slices with opposite momenta $\bm{k}$ and $-\bm{k}$ together with their winding invariants are paired up under $\mathcal{T}$, and the vortex lines induce jumps of the windings by even integers. 
%The story up to this point of \emph{CI semimetals}~\cite{Chiu2016} applies equally well to electrons and classical waves, provided $\mathcal{S}$ and $\mathcal{T}$ are symmetries. However, the dynamical structure for MSSWs is fundamentally different from that of electrons: the time-evolution operator for dipolar spin waves belongs to a \emph{symplectic group}~\cite{Arnold1989} instead of a unitary group. Moreover, the symplectic two-form plays the role of the chiral symmetry $\mathcal{S}$, upon a choice of metric. The upshot is that the eigenvalues of the Hamiltonian with this $\mathcal{S}$-symmetry are directly related to although they are directly related for classical eigenfrequencies, even though they do not coincide in general. Both $\mathcal{S}$ and $\mathcal{T}$ are intricately intertwined with the classical mechanical nature of the problem as the latter embodies the reality condition of classical systems. The symplectic structure, $\mathcal{T}$, and $\mathcal{S}$ are essential in explaining why MSSWs are not gapless, unidirectional, and robust only for propagations perpendicular to the magnetization.

\begin{figure}[tbp]
\includegraphics[width=0.48\linewidth]{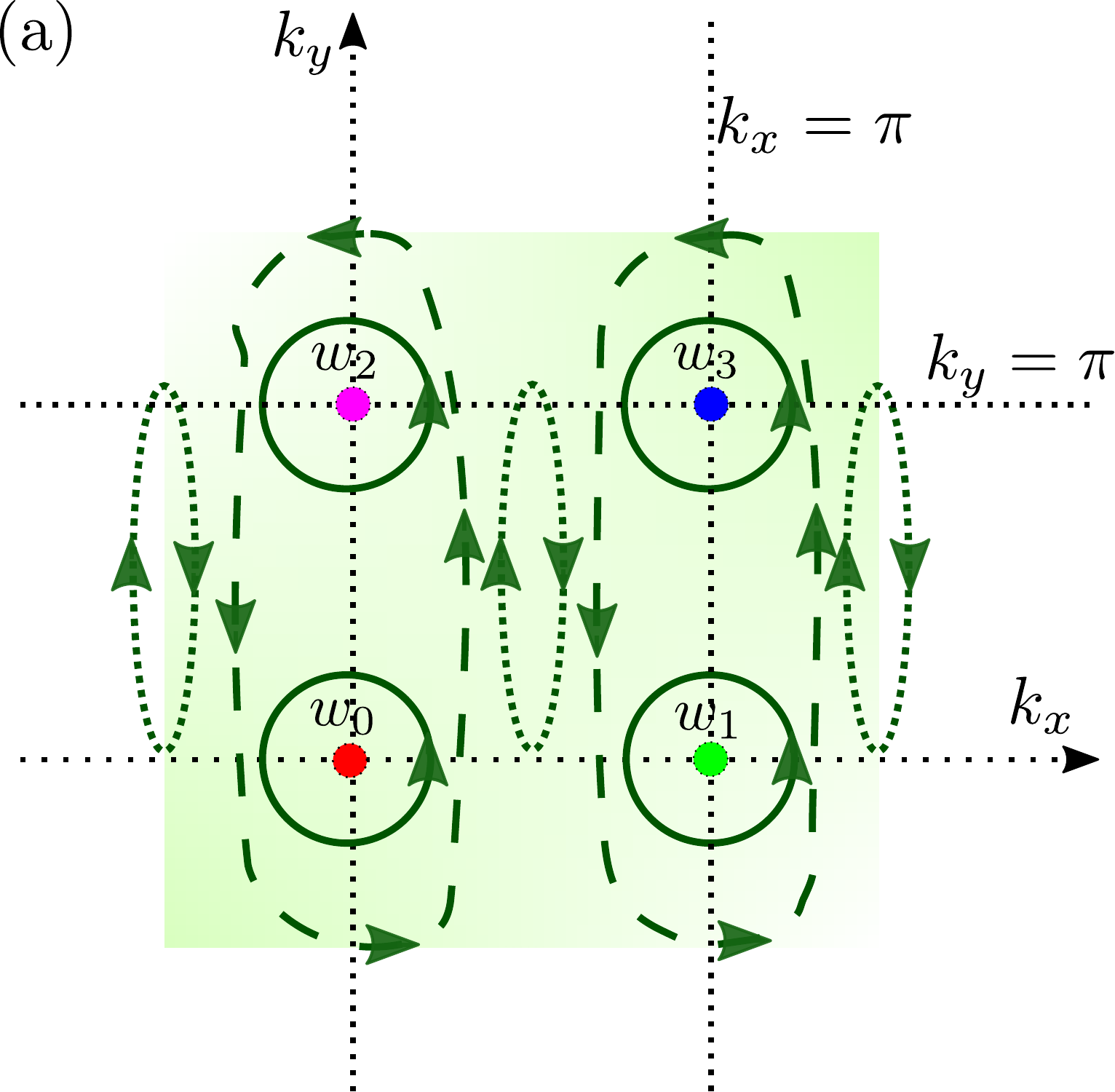}
\includegraphics[width=0.48\linewidth]{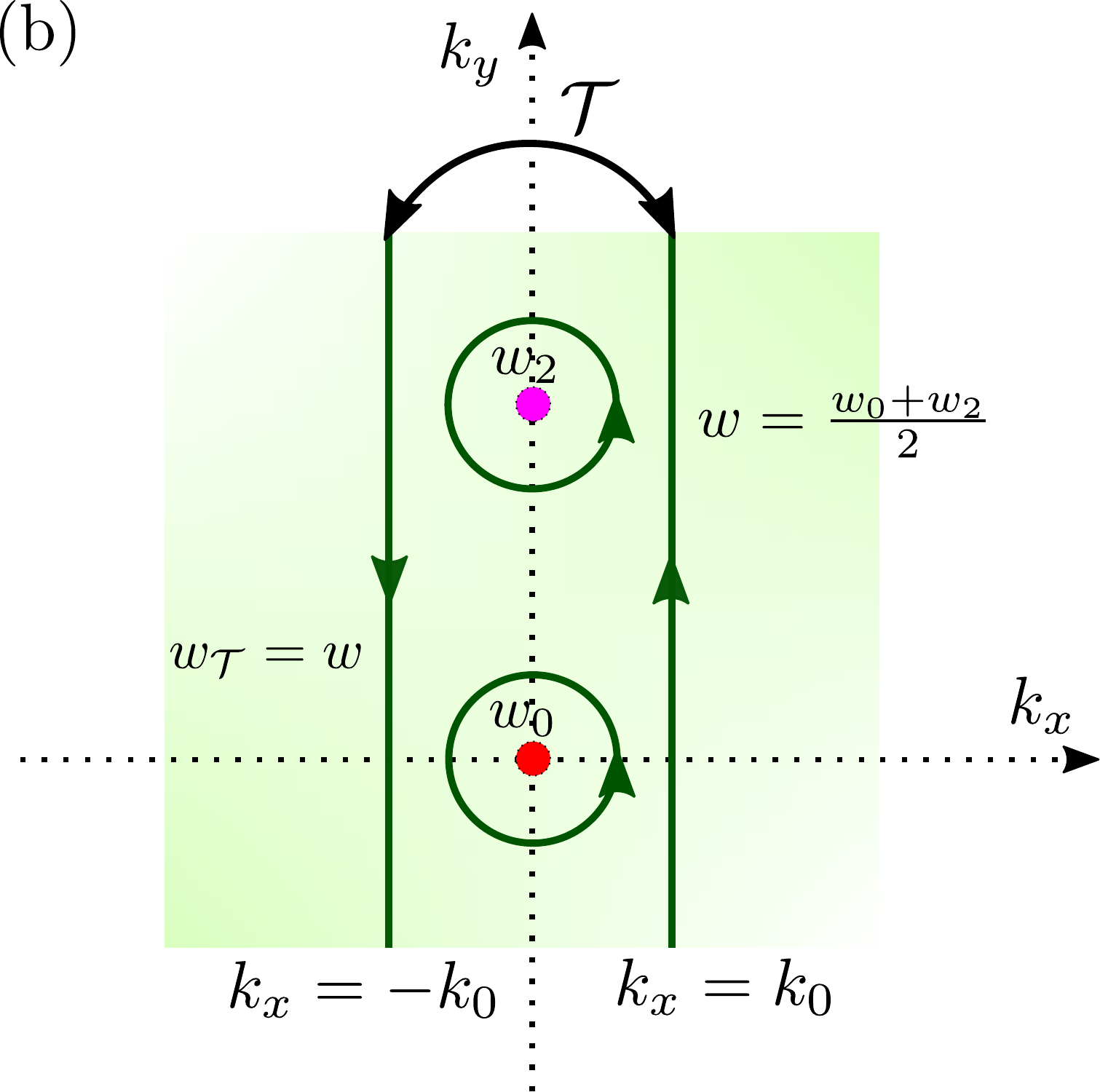}
\caption{(a) Graphical proof of the charge cancellation. Due to $2\pi $ periodicity in $k_x ,k_y $, the solid, dashed and dotted loops are continuously deformable to each other without crossing TRIMs. The dotted loops are contractible to two points, hence trivial with zero winding. (b) Determination of the winding of large loops by a deformation of small loops. Due to $\mathcal{T}$-symmetry, large loops $k_x = k_0 $ and $k_x = -k_0 $ have the same windings $w =w_{\mathcal{T}}$.
% However, their orientation with respect to that of $k_y $ is opposite to each other. 
}
\label{fig:kspace}
\end{figure}

We elaborate on this structure by elementary winding number analysis augmented with $\mathcal{T}$-symmetry, following ideas in Refs.~\cite{Mathai2017,Thiang2017}. We assume that the system is periodic on a 3D lattice $\mathbb{Z}^3$ and denote the Brillouin zone by $\mathbb{T}^3$. By definition of a class CI Hamiltonian $H$~\cite{Ryu2010}, given are a unitary $\Gamma $ and an antiunitary $\mathcal{T}$ such that $\{ H ,\Gamma \} = [ H ,\mathcal{T} ] = \{ \Gamma ,\mathcal{T} \} = 0$, $\Gamma ^2 = \mathcal{T}^2 = 1$  where $[ \cdot ,\cdot ]$ ($\{ \cdot ,\cdot \} $) denotes (anti-)commutator. In the Brillouin zone, $\Gamma $-symmetry means
\begin{equation}
H_{\bm{k}} = \begin{pmatrix}
	0 & U_{\bm{k}} \\
	U^{\dagger }_{\bm{k}} & 0 \\
	\end{pmatrix},\qquad \bm{k}=(k_x,k_y,k_z)\in\mathbb{T}^3, \label{eq:chiral_Hamiltonian}
\end{equation}
in a basis in which $\Gamma =\mathbbm{1} \otimes \sigma _3$ ($\sigma _{1,2,3}$ denote Pauli matrices), while time-reversal symmetry $\mathcal{T}$ relates the Hamiltonian at $\bm{k}$ and $-\bm{k}$ by $U_{-\bm{k}} = U^t_{\bm{k}}$. Suppose $H_{\bm{k}}$ is {\it gapped}, i.e.\ its eigenvalues are all nonzero, on $\mathbb{T}^3 \backslash L $ where $L = \{ \bm{k} \in \mathbb{T}^3 | k_x = 0,\pi , k_y =0,\pi \} $ is a set of four vortex lines parallel to $k_z$. More general line defects are obtained by either deforming or splitting the four straight lines passing through the TRIMs on $k_z =0$ plane~\footnote{See Supplemental Material for general vortex line configurations, details of the classical eigenfrequency problem, explicit edge state solutions, and additional information on the Fourier transform.}. Here we focus on the straight line configuration realized by MSSWs for readability. The gap condition means $\det U_{\bm{k}} \neq 0 $ on $\mathbb{T}^3 \backslash L$. Let us first examine the slice $\mathbb{T}^2 = \{ \bm{k} \in \mathbb{T}^3 | k_z = 0 \} $, which the vortex lines intersect at its four time-reversal invariant momenta (TRIMs). Take a small but otherwise arbitrary loop $\ell _a $ encircling only the $a$-th TRIM (labelling in FIG.~\ref{fig:kspace}(a)), oriented counterclockwise. Define its winding number by
\begin{equation}
w_a = \frac{1}{2\pi i}\oint _{\ell _a } d  \left\{ \ln \left( \det U_{\bm{k}} \right) \right\}  , \quad a =0,1,2,3 .\label{eq:winding}
\end{equation}
The winding number is an integer \emph{topological invariant}, insensitive to perturbations of $U$ (thus of $H$ that respects $\Gamma $ and the gap condition), and deformations of $\ell_a$ (avoiding the vortices). As graphically proven in FIG.~\ref{fig:kspace}(a), there is a ``charge cancellation'' consistency condition $\sum _{a=0}^3 w_a = 0$ because the sum may be evaluated in a second way which is manifestly trivial. 

By $\mathcal{T}$-symmetry, the $w_a$ are enough to determine the winding numbers along ``large" loops of $\mathbb{T}^2$ (say, at constant $k_x$ or $k_y$, FIG.~\ref{fig:kspace}(b)). First, any small loop $\ell_a$ can be deformed into a symmetric one which is mapped onto itself under $\mathcal{T}: \bm{k} \rightarrow -\bm{k}$. In Eq.~(\ref{eq:winding}), the integrand for one half of $\ell_a$ is repeated on the other half, so that the total line integral should be an \emph{even} integer. Similarly, a large winding around $k_y$ at a fixed $k_x =k_0 \neq 0,\pi$ must equal that at $k_x =-k_0$ evaluated along the opposite orientation, and they are constrained by their sum equalling that of the enclosed small windings. As for the 2D slices with $k_z \neq 0,\pi $ which do not respect $\mathcal{T}$ individually, continuity along $k_z$ forces on them the same topological structure as the $k_z =0 $ slice (FIG.~\ref{fig:geometry}(b)). To summarize, Hamiltonians in class CI with 1D line defects $L$ are topologically characterized
%~\footnote{Mathematically, a cohomology calculation verifies $H^1_{\mathbb{Z}_2}\left( \mathbb{T}^3 \setminus L \right)  = \left( 2\mathbb{Z}\right) ^3 $, where the $\mathbb{Z}_2$ subscript denotes $\bm{k}\mapsto -\bm{k}$. Via the line integral (\ref{eq:winding}), these classes pair with homology classes represented by small loops $\ell_a$ to give even integers.} 
by three independent small even windings $(w_0,w_1,w_2)\in(2\mathbb{Z})^3$. If $w_a \neq 0$, there is a corresponding vortex line of topologically protected gapless points or singularities of $H_{\bm{k}}$.

To obtain the \emph{CI semimetal bulk-edge correspondence}, consider for some fixed $k_x,k_z \neq 0,\pi $, the two class AIII 1D subsystems $\mathbb{T}_{(k_x,k_z)}$ and $\mathbb{T}_{(-k_x,-k_z)}$ along the $y$ direction. Their (large) winding numbers are equal and opposite by $\mathcal{T}$-symmetry, and if nonzero, the 1D bulk-edge correspondence of class AIII ensures that when a surface is cut along $x$-$z$ plane, there appear surface-localized eigenstates of $H$ with zero eigenvalue. A similar argument holds with $x$ replaced by $y$. If at least one $w_a$ is nonzero, then some $\mathbb{T}_{k_x ,k_z} $ or $ \mathbb{T}_{k_y , k_z} $ has nonzero winding number, implying the existence of edge eigenstates.

The application of the CI semimetal setup presented above requires a Hamiltonian \emph{operator} acting on a complex Hilbert space. To introduce such a structure for classical mechanical systems on a lattice $\mathbb{Z}^3$, a metric plays a crucial role; below we explain why~\cite{Note1}. %~\footnote{See Supplemental Material for a presentation in terms of tensor components}. 
In classical mechanics~\cite{Arnold1989}, one starts from a real symplectic vector space $V$ whose coordinates are canonical variables $v = (\{ p_{\bm{n}} \} ,\{ q_{\bm{n}}\} )^t \in V , \bm{n} \in \mathbb{Z}^3 $. The symplectic two-form $\omega = \sum _{\bm{n}} dp_{\bm{n}} \wedge dq_{\bm{n}}$ can be regarded as a linear map identifying $V$ with the dual space $V^{\ast }$. In linearized problems, the dynamics is determined by a positive definite quadratic energy function $E(v)$, i.e.\ another linear map $V \rightarrow V^{\ast }$. Hamilton's equations of motion read $dv/dt = I\circ E( v) $, where $I = \omega ^{-1}:V^{\ast }\rightarrow V$ is the Poisson bracket and $\circ $ denotes composition of maps. \emph{Note that $E$ is not an operator} (a map $V\rightarrow V$). One way of promoting the energy $E:V\rightarrow V^{\ast }$ to an operator is to assume that a preferred metric $g:V\rightarrow V^{\ast }$ is given on $V$ and define $H = g^{-1}\circ E : V\rightarrow V$. Indeed $H$ defined in this way is what one calls Hamiltonian in problems where $V$ comes with a natural Euclidean metric. Now the equations of motion may be rewritten as $dv/dt = JHv$ where $ J =I\circ g : V \rightarrow V $ satisfies $J^t =-J$ (transpose with respect to $g$). By rescaling $g^{\prime } = g \circ \sqrt{-( I\circ g ) ^{-2}}$, we can further arrange for $J^2 =-1$. 

%When carrying out Fourier transforms, one complexifies $V$ to $V^{\mathbb{C}} = V \oplus iV$, and extends $g,H,J$ complex linearly to $V^{\mathbb{C}}$. We then speak of symmetry classes of $H$ in parallel to quantum mechanical electrons; here, $H$ always has the even"time-reversal symmetry" $\mathcal{T}$ of complex conjugation, reflecting the reality of the original problem. 

%Having specified the notion of classical mechanical Hamiltonians, we now discuss their symmetries and topology. 
For a given classical system with a (rescaled) metric as above, i.e.\ maps $\omega ,E, g:V\rightarrow V^{\ast }$, we shall say $H = g^{-1}\circ E $ belongs to \emph{class CI*} if there exists a positive $h_0 \in \mathbb{R}^+ $ such that $\{ H-h_0 ,J\} =0$ with $J =\omega ^{-1}\circ g$. To recognize the connection to the definition of class CI, we complexify $V$ to $V^{\mathbb{C}} = V\oplus iV$ and extend $g,H,J$ complex linearly to $V^{\mathbb{C}}$. This step is usually implicit when one carries out Fourier transforms. Here, $H$ always has the even ``time-reversal" symmetry $\mathcal{T}$ of complex conjugation, reflecting the reality of the original problem. One can also introduce a chiral symmetry $\Gamma  =iJ$, which is unitary and satisfies $\{ H -h_0,\Gamma  \} = \{ \mathcal{T} ,\Gamma \} =0$, $\Gamma ^2 =1$. Therefore, a classical $H$ in class CI* has the complexified $H-h_0$ in class CI. If $H-h_0 $ is semimetallic with vortex lines $L$, the winding numbers $(w_0 ,w_1 , w_2 )$ topologically characterize $H$. In a basis where $\Gamma = \mathbbm{1} \otimes \sigma _2 $, the classical Hamiltonian takes its canonical form
\begin{equation}
H = h_0 + S\otimes \sigma _1 + C\otimes \sigma _3  \label{eq:chiral_Hamiltonian2}
\end{equation}
with some \emph{real} operators $S,C$. A basis transform by $Q = \{ 1+i(\sigma _1 + \sigma _2 + \sigma _3 )\} /2$ brings $\Gamma  $ into $Q^{\dagger }\Gamma Q = \mathbbm{1}\otimes \sigma _3 $ and $H -h_0 $ into the off-diagonal form as in Eq.~(\ref{eq:chiral_Hamiltonian}) with the Fourier transform of $U= C -iS $ providing the winding numbers, Eq.~(\ref{eq:winding}). If some $w_a \neq 0$, the CI semimetal bulk-edge correspondence predicts edge states in the gap of $H$ at $h_0$, i.e.\ $Hv_{n_0} = h_0 v_{n_0}$.

We now reveal that the edge states $v_{n_0}$ appear \emph{above the physical bulk frequency spectrum}. Although the eigenvalues of $H$ do not equal physical eigenfrequencies in general, there is a one-to-one correspondence between them within class CI*. Suppose $0\neq v_{n+} \in V $ is an eigenvector of $H$ with eigenvalue $h_0 +\epsilon _n >0$. The class CI* condition $\{ H-h_0 ,J \} =0$ implies that $v_{n-} \equiv Jv_{n+} \in V$ satisfies $Hv_{n-} = (h_0 -\epsilon _n ) v_{n-} $. Whether $\epsilon _n =0$ or not, $v_{n-} \not\propto v_{n+} $ because the eigenvalues of $J$ are $\pm i$ while $v_{n\pm }$ are both real. Hence all eigenvectors of $H$ come in pairs $v_{n\pm }$ mutually related by $J$ with respective eigenvalues $h_0 \pm \epsilon _n $. One can choose the label $n$ such that $\epsilon _n \geq 0$. Since $v_{n\pm }$ form a complete set of basis vectors, the general solution of Hamilton's equations $dv/dt =JHv$ is given by $v = \sum _{n,\pm }c_{n\pm }(t) v_{n\pm }$ with the time-dependent coefficients satisfying 
\begin{equation}
\frac{d}{dt}\begin{pmatrix}
	c_{n+}\\
	c_{n-} \\
	\end{pmatrix} = \begin{pmatrix}
	0 & -h_0 + \epsilon _n \\
	h_0 + \epsilon _n & 0 \\
	\end{pmatrix} \begin{pmatrix}
	c_{n+} \\
	c_{n-} \\
	\end{pmatrix} .
\end{equation}
This yields
\begin{equation}
c_{n\pm } =A_n (h_0 \pm \epsilon _n )^{-1/2}\cos (\omega _n t +\alpha _n \mp \pi /4) \label{eq:solution}
\end{equation}
where $A_n ,\alpha _n $ are constants and $\omega _n = \sqrt{h_0^2 -\epsilon _n^2 }$ is the physical eigenfrequency. This clearly shows that the edge states with $\epsilon _{n_0}=0$ have the physical frequency $\omega _{n_0} =h_0 $ higher than those of the bulk modes with $\epsilon _n \neq 0$.

While our topological characterization of class CI Hamiltonians is interpreted in the classical mechanical framework, previous studies of topological spin waves~\cite{Shindou2013,Peano2018,Lu2018} focused on eigenvalues of $iJH$ in the Bogoliubov-de Gennes formalism. To the best of our knowledge, their approach seems to always predict gapless edge modes, and consequently Ref.~\cite{Shindou2013} missed the topological nature of MSSWs.

%\footnote{Another way is to complexify the problem first and take $V^{\mathbb{C}}$ with an indefinite metric $i \omega $ as a Krein space. This path has been taken in \cite{Peano2018} (also see \cite{Lu2018}).}. 
%%%%%
%%%%% Is the Peano approach known to be really different?
%%%%%
%One then usually complexifies $V$ by $V^{\mathbb{C}} = V \oplus iV$, which is often implicitly done when Fourier transforms are used. The metric and operators $H,J$ are extended complex linearly to $V^{\mathbb{C}}$, and one can speak of symmetry classes of $H$ in parallel to quantum mechanical electrons. In particular, a classical $H$ always has an even time-reversal $\mathcal{T}$ that is the complex conjugation, which simply embodies the reality of the original problem. 

%o relate them to $H$, we shall say that $J$ acts as a chiral symmetry if  $\{ H -h_0 ,J \} = 0 $. This ensures that $\Omega = \sqrt{h_0^2 - ( H-h_0 )^2 }$, and hence the eigenvectors of $H$ and $\Omega $ clearly coincide and the eigenvalues $\pm \epsilon _n $ of $H-h_0 $ are related to those of $\Omega $ by $\omega _n = \sqrt{h_0^2 -\epsilon _n^2 }$. 

%Moreover, the middle of ``gap" of $H$ in this setup is $\epsilon _{n_0} =0$ and the corresponding eigenfrequency at which ``gapless edge modes" would appear is greater than all the others $\omega _{n_0} = h_0 > \omega _n , n \neq n_0$. 
%%%%%
%%%%% Should show the basis transformation that turns $\mathcal{S} = \mathbbm{1} \otimes \sigma _3 $ and similarly changes $\mathcal{T}$ to that in the earlier part of the paper.
%%%%%

To summarize, classical problems with a metric have a natural candidate for chiral symmetry in $ \Gamma = iJ=i \omega ^{-1}\circ g $. If $H$ up to a constant shift anticommutes with $\Gamma $, the (real) eigenvectors of $H$ do coincide with the physical eigenstates, while its eigenvalues $h_0 \pm \epsilon _n $ correspond to the physical eigenfrequencies $\omega _n = \sqrt{h_0^2 -\epsilon _n^2}$. If there is a ``gapless" edge ``state" of $H$ ($\epsilon _{n_0}=0$) protected by a CI semimetal structure, there exists an edge-localized physical eigenstate whose frequency ($\omega _{n_0} =h_0$) appears above the bulk frequency spectrum.

\begin{figure}[tbp]
\includegraphics[width=\linewidth]{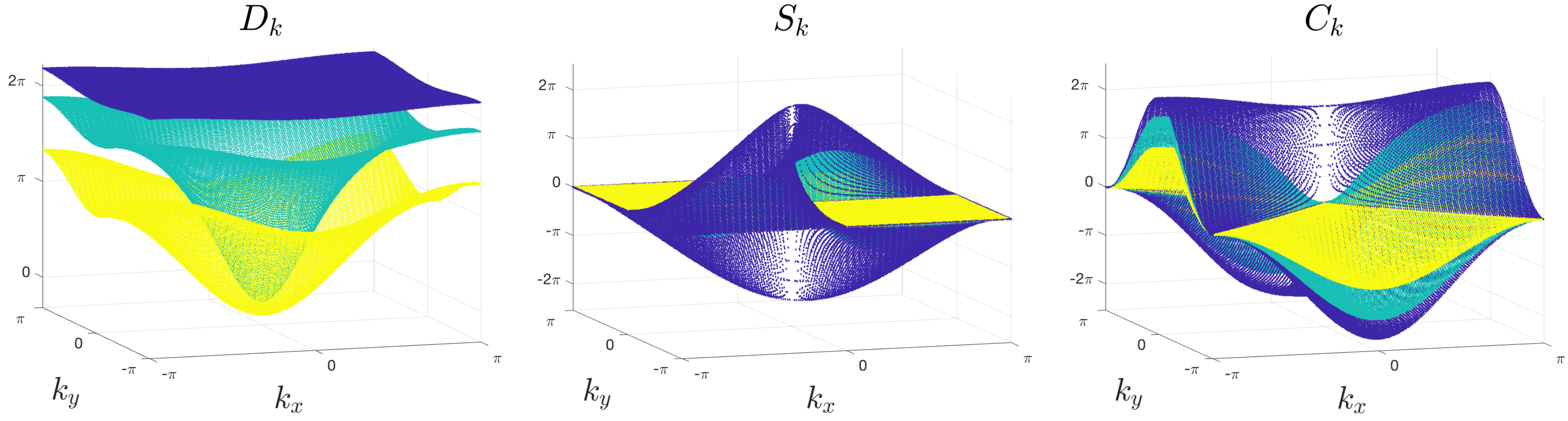}
\caption{Numerically evaluated Fourier transform of the Hamiltonian (\ref{eq:dipolar_Hamiltonian}) for constant $k_z $ slices (we set $G=1$): $D_{\bm{k}}$ (Left), $S_{\bm{k}}$ (Center) and $C_{\bm{k}}$ (Right) with $k_z = 0 $ (blue), $0.2\pi $ (green) and $0.5\pi $ (yellow).}
\label{fig:Fourier}
\end{figure}

The general framework presented above requires only the specified symmetry conditions. We now demonstrate that all those assumptions are almost faithfully respected by dipolar spin waves traveling perpendicular to the magnetization~\cite{Damon1961}. Consider a simple cubic lattice of classical spins interacting only with an external magnetic field $B>0$ along $z$ direction and between each other via dipole-dipole interactions. The ground state satisfies $\bm{s}_{\bm{n}} = (0,0,1) , \bm{n}\in \mathbb{Z}^3 $ where $\bm{s}_{\bm{n}}$ is the normalized spin vector at site $\bm{n}$. The energy function of spin waves in terms of the linearized spin components $\bm{s}_n \approx  (s^x_{\bm{n}},s^y_{\bm{n}},1 - \{ (s^{x}_{\bm{n}})^2 + (s^{y}_{\bm{n}})^2 \}/2)$ yields~\cite{Aharony1973}
\begin{equation}
E = B^{\prime }\sum _{\bm{n}} s^{\alpha }_{\bm{n}}s^{\alpha }_{\bm{n}} - G \sum _{\bm{n}\neq \bm{n}^{\prime }} \frac{\partial ^2}{\partial n^{\alpha } \partial n^{\prime \beta }}\left( \frac{1}{\left| \bm{n}-\bm{n}^{\prime } \right| } \right) s^{\alpha }_{\bm{n}}s^{\beta }_{\bm{n}^{\prime }} \label{eq:dipolar_Hamiltonian}
\end{equation}
where sums over $\alpha ,\beta =x,y$ are implicit, $B^{\prime } = B+4\pi G/3$~\footnote{This expression of $B^{\prime }$ is due to Ref.~\cite{Holstein1940}. Also see Supplemental Material.} %For an infinite cubic lattice, it should have been $B^{\prime }=B$~\cite{Cohen1955}. However, the former agrees with the continuum limit used in Ref.~\cite{Damon1961} while the latter result relies on the exact cubic symmetry.The value of $B^{\prime }$ does not affect the topological nature of MSSWs.
and the constants $B$ and $G$ are appropriately normalized. $s^x_{\bm{n}}$ and $s^y_{\bm{n}}$ are identified to be $p_{\bm{n}},q_{\bm{n}}$ respectively with the area two-form of the sphere (phase space of $\bm{s}_n$) acting as the symplectic two-form $\omega $~\cite{Stancil2009}. The system comes with the Euclidean metric $g= \delta _{\alpha \beta }$ of the spin configuration space, with which the Hamiltonian $H$ is identical to $E$ as a matrix. Applying spatial Fourier transform, $H$ decomposes over the Brillouin zone as two-by-two matrices $H_{\bm{k}} =\left( B + D_{\bm{k}} \right) \mathbbm{1}+ S_{\bm{k}} \sigma _1 + C_{\bm{k}} \sigma _3 $ ($\mathbbm{1}$ is the unit matrix) each acting on $(p_{\bm{k}} ,q_{\bm{k}})$. For $\bm{k} \approx 0$, i.e.\ in the long-range limit, the coefficient functions are approximated by
\begin{equation}
D_{\bm{k}}  = 2\pi G \frac{k_x^2 +k_y^2}{\left| \bm{k} \right| ^2} , S_{\bm{k}} = 2\pi G \frac{2k_x k_y}{\left| \bm{k} \right| ^2} , C_{\bm{k}} = 2\pi G \frac{k_x^2 -k_y^2}{\left| \bm{k} \right| ^2} .
\end{equation}
$H_{\bm{k}}$ is already in the class CI* canonical form Eq.~(\ref{eq:chiral_Hamiltonian2}) with $h_0 =B+D_{\bm{k}}$ and has complex conjugation as a $\mathcal{T}$-symmetry. $\sigma _2$ is identified with a chiral symmetry, which is exact when $D_{\bm{k}}$ is constant. To compute the winding number, note $U_{\bm{k}} = C_{\bm{k}} -iS_{\bm{k}}$ as stated below Eq.~(\ref{eq:chiral_Hamiltonian2}). Substituting it into Eq.~(\ref{eq:winding}) yields $w_0 =-2$ around the origin (and $k_z$ axis), \emph{proving that the dipolar Hamiltonian is topologically non-trivial}. Although expressions for $D_{\bm{k}},S_{\bm{k}},C_{\bm{k}}$ away from the origin are not available in a closed analytical form, they can be numerically evaluated by Ewald's method~\cite{Cohen1955} as plotted in FIG.~\ref{fig:Fourier}. One confirms $U_{\bm{k}}\neq 0 $ along $\bm{k} =(0,\pi ,k_z ) , (\pi ,0,k_z )$ and $U_{\bm{k}} =0 $ (i.e.\ a vortex line is located) along $\bm{k} = (\pi ,\pi ,k_z )$~\cite{Note1}. %~\footnote{See Supplemental Material for computational details of the numerical Fourier transform.}. 
Thus the topology of the dipolar spin wave Hamiltonian is characterized by $(w_0 ,w_1 ,w_2 ) = (-2,0,0 )$. All the 1D slices for fixed $k_x,k_z \neq 0,\pi $ have winding numbers $\pm 1$. Note that the slices $\pm (k_x ,k_z )$ are paired by the reality condition (``$\mathcal{T}$-symmetry'') and represent the same physical degrees of freedom. Therefore, when a surface is cut along $x$-$z$ plane, one surface mode for each $k_x ,k_z $ is expected. 

Strictly speaking, the bulk-edge correspondence is valid only if $D_{\bm{k}}$ is constant. It is satisfied on the $k_z =0$ slice in the long-range limit as $D_{\bm{k}}\rightarrow 2\pi G$ and the eigenfrequency of the corresponding edge mode should be $\omega = B+2\pi G$, which is precisely the frequency of MSSWs for $k_z =0$.
%~\footnote{The prediction $\omega =B+2\pi G ={\rm const}$ for MSSWs applies to the case of a half-infinite space. The deviation from this relation in Ref.~\cite{Damon1961} is due to the interaction between the edge modes localized on the opposite surfaces.}
Although $D_{\bm{k}}$ deviates from $2\pi G$ for $k_{x,y}$ of order unity, the numerical calculation shows the $\bm{k}$ dependence is weak so that the chiral symmetry is approximately satisfied for $k_z =0$ (FIG.~\ref{fig:Fourier}). In contrast, on planes with constant $k_z \neq 0$, $D_{\bm{k}}$ varies as much as $S_{\bm{k}}$ or $C_{\bm{k}}$ does and chiral symmetry is violated. This can explain the lack of robustness of obliquely traveling MSSWs. Physically, we expect the chiral symmetry breaking term $D_{\bm{k}}$ to shift the frequency of the edge modes relative to that of bulk modes, eventually causing them to merge with the bulk band and disappear. To our knowledge, the fate of the class AIII bulk-edge correspondence when strict chiral symmetry is broken while the bulk winding is still well-defined is an open mathematical problem.
% and we cannot be conclusive about this point at present. 

Finally we discuss the chiral, unidirectional propagation of MSSWs. When a surface is cut in the $y$ direction as in Fig.~\ref{fig:geometry}(b), edge states appear on the surface Brillouin zone except for the projections of the bulk vortex lines $k_x =0,\pi $. Thus the edge states always have nonzero components of $k_x$ and one can define their chirality with respect to the $x$ direction. The reality condition $\mathcal{T}$ means the pair of edge states at $\pm (k_x ,k_z )$ are physically identical so that the sign of $k_x$ itself cannot decide the direction of propagation. This however also implies there is one propagating mode for the pair of states, which is thus necessarily chiral (i.e.\ it can propagate in only one of $\pm (k_x ,k_z) $ directions). The ``chiral symmetry" $\Gamma $  is indeed correlated with the direction of propagation in the following way. Recall that class AIII edge states are eigenstates of $\Gamma $ with their eigenvalues $s = \pm 1$ for windings $\pm 1$~\cite{Ryu2002,Gomi2018,Graf2018}. Due to the $\mathcal{T}$-symmetry, edge states with $s =\pm 1$ are paired up and form a single physical eigenstate. An explicit computation~\cite{Note1} %~\footnote{See Supplemental Material for an explicit construction of chiral edge state solutions.}, 
shows that $s= +1 $ for $k_x \gtrless 0 $ gives edge modes traveling in the positive and negative $x$ directions respectively. 

%The sign of the phase velocity is essentially governed by the ``signature" of $\hat{J}\Omega $. It can be unambiguously defined for MSSWs if we recall that the edge modes of class AIII are eigenstates of the chiral symmetry operator $\mathcal{S}$~\cite{}. This implies that $J $ and $\hat{J}$ restricted on the subspace $V_0 \subset V $ spanned by the edge eigenvectors $v_{n_0 \pm } \in V$ are identical up to an orthogonal transformation. From the skew symmetry of $J,\hat{J}$, it follows either $\hat{J} = +J$ or $\hat{J} =-J$ on $V_0$ and the corresponding chirality of MSSWs is positive or negative respectively. 

In conclusion, we have established the notion of class CI semimetals characterized by even windings around vortex defect lines, and explained how they arise in certain classical mechanical systems. We constructed a chiral symmetry operator from the symplectic two-form and a metric. We showed that the corresponding chiral symmetric classical systems can support topologically protected edge modes with their eigenfrequencies appearing above the bulk spectrum. The framework is applicable to MSSWs for $k_z =0$ and reproduces all of their characteristic features. 

\begin{acknowledgements}
The authors would like to thank Shunsuke Daimon, Kiyonori Gomi, Kazuya Harii, Jun'ichi Ieda, Max Lein, Ben McKeever, Michiyasu Mori, Naoto Nagaosa, Koji Sato and Libor \v{S}mejkal for helpful comments. This work was supported by the Transregional Collaborative Research Center (SFB/TRR) 173 SPIN+X of the DFG, JSPS KAKENHI Grant Number JP 18H05855, Australian Research Council DE170100149, the German Research Foundation (DFG) No. EV 196/2-1 and No. SI 1720/2-1, the KIST Institutional Program, JST-ERATO `Spin Quantum Rectification' and AIMR Tohoku University.
\end{acknowledgements}

\bibliographystyle{apsrev4-1}
\bibliography{bibliography_MSSWs}

\end{document}